\documentclass[a4paper,11pt]{article}
\usepackage{latexsym}
\usepackage{epsfig, subfigure, subfloat, graphicx, float }
\usepackage{amssymb}
\usepackage{amsmath}
\usepackage{epstopdf}
\usepackage{authblk}
\title{Symmetron and de Sitter attractor in a teleparallel model of cosmology}

\author[1]{H. Mohseni Sadjadi\thanks{mohsenisad@ut.ac.ir}}

\affil[1]{Department of Physics, University of Tehran}

\begin{document}

\maketitle

\begin{abstract}
In the teleparallel framework of cosmology, a quintessence with non-minimal couplings to the scalar torsion and a boundary term is considered. A conformal coupling to matter
density is also taken into account. It is shown that the model can describe the onset of cosmic acceleration after an epoch of matter dominated era, where dark energy is negligible, via $Z_2$ symmetry breaking. While the conformal coupling holds the Universe in a state with zero dark energy density in the early epoch, the non-minimal couplings lead the Universe to a stable state with de Sitter expansion at the late time.
\end{abstract}
\section{Introduction}
In teleparallel model of gravity, instead of torsion-less Levi-Civita, curvature-less Weitzenb\"{o}ck connections are used \cite{tele,tele1}.
In this model the action reads
\begin{equation}\label{1}
S=\int \left[-{T\over 2k^2}+\mathcal{L}_m\right]\sqrt{-g}d^4x.
\end{equation}
The matter Lagrangian density is denoted with $\mathcal{L}_m$. In terms of $G_N$ which denotes the Newtonian gravitational constant, $k^2$ is given by $k^2=8\pi G_N$.
The torsion scalar, $T$, is
\begin{equation}\label{2}
T={1\over 4}{T^\rho}_{\mu \nu}{T_\rho}^{\mu \nu}
+{1\over 2}{T^\rho}_{\mu \nu}{T^{\nu \mu}}_{\rho}-{T^\rho}_{\mu \rho}{T^{\nu \mu}}_{\nu},
\end{equation}
where ${T^a}_{\mu \nu}=\partial_\mu e^{a}_{\nu}-\partial_\nu e^{a}_\mu$ is the torsion tensor. Vierbeins are denoted with ${e_a}^\mu$, whose duals are ${e^a}_\mu$.  The metric tensor is $g^{\mu \nu}=\eta_{ab}{e_a}^\mu {e_b}^\nu$ and $e=det({e^a}_\mu) =det\sqrt{-g}$ (for more details bout teleprallel gravity see \cite{tele,tele1}) .

The sum of the Ricci scalar, $R$,  and the torsion scalar, $T$,  is a total divergence \cite{baha,baha1}
\begin{equation}\label{3}
R+T=2\nabla_\mu T^\mu,
\end{equation}
where $T_\mu={T^\nu}_{\nu\mu}$.  Therefore we find that the equations of motion derived from (\ref{1}) are  equivalent to those derived from Einstein-Hilbert action
\begin{equation}\label{4}
S=\int \left[{R\over 2k^2}+\mathcal{L}_m\right]\sqrt{-g}d^4x.
\end{equation}
One way to explain the inflationary regime of the early Universe and also the late time acceleration is to introduce exotic scalar field in the Einstein-Hilbert action. If this scalar field is minimally coupled to gravity, we find that the difference between (\ref{1}) and (\ref{4}) becomes again a total divergence and  both theories give the same equations of motion, but for non-minimal coupling this situation changes. Let us consider an action with non-minimal coupling term $\epsilon R \phi^2$ \cite{nm1,nm1.1},
\begin{equation}\label{5}
S=\int \left[{R\over 2k^2}+{1\over 2}\left(\partial_\mu \phi \partial^\mu \phi +\epsilon R \phi^2\right)-V(\phi)+\mathcal{L}_m\right]\sqrt{-g}d^4x.
\end{equation}
This non-minimal coupling is required for renormalizability of the theory, and has also root in quantum corrections to scalar field theory on curved space-time \cite{nm1}. This model has been employed to study the inflationary epoch by attributing the  Higgs scalar to the inflaton \cite{nm1.1}.  Inspired by (\ref{5}), the action
\begin{equation}\label{6}
S=\int \left[-{T\over 2k^2}+{1\over 2}\left(\partial_\mu \phi \partial^\mu \phi -\epsilon T \phi^2\right)-V(\phi)+\mathcal{L}_m\right]\sqrt{-g}d^4x
\end{equation}
has been introduced to describe the positive acceleration of the Universe as well as the possible phantom divide line crossing \cite{tele1}. Due to the presence of the nonminimal coupling, i.e. $\epsilon\neq 0$,  (\ref{5}) is not equivalent to (\ref{6}). In \cite{baha} a more general action was proposed
\begin{equation}\label{7}
S=\int \left[-{T\over 2k^2}+{1\over 2}\left(\partial_\mu \phi \partial^\mu \phi -\epsilon T \phi^2-\chi B \phi^2\right)-V(\phi)+\mathcal{L}_m\right]\sqrt{-g}d^4x,
\end{equation}
where $B=2\nabla_\mu T^\mu$. This model includes (\ref{5}), and (\ref{6}) as its subclasses. This can be seen by setting  $\epsilon=-\chi$, and  $\chi=0$.
Recently, different cosmological and gravitational aspects of this model were discussed in the literature \cite{baha2}.

Teleparallel models of gravity, such as the scalar-torsion model,  suffer from the lack of invariance under local Lorentz transformations of the tetrads \cite{soti}. This noninvariance occurs also in the model (\ref{7}) \cite{baha}. To solve this problem, in \cite{tama}, the possibility to find “good tetrads” was discussed. The preferred-tetrad-frame cannot be
detected by only  measuring the components of the metric. Such measurements
will employ some gauge fields to restore the local Lorentz symmetry, giving rise to a Poincar\'{e} gauge theory \cite{hay}. Recently it was proposed that by employing purely inertial spin connection and flat tetrad in a covariant manner, the Lorentz invariance may be held \cite{kr}. In this framework,  there is no additional spin contribution in the cosmological equations, and so we will expect that the  model (\ref{7}), like scalar-torsion gravity, is still valid in this covariant formalism \cite{jar}.

In this paper, we consider (\ref{7}) in a spatially flat Friedmann-Lema\^{i}tre-Robertson-Walker (FLRW) space-time filled nearly with pressureless matter and dark energy. Based on this action we try to explain the late time cosmic acceleration after an epoch of matter domination.  To do so,  besides the non-minimal couplings, we also need to a conformal coupling between the quintessence and pressureless matter. This kind of coupling has been previously considered in screening model, such as chameleon \cite{cham} and symmetron \cite{sym}. In the symmetron model, cosmic acceleration is related to a $Z_2$ symmetry breaking triggered  by reduction of dark matter density in the late time. However, due to the large mass of quintessence the scalar field overshoots rapidly and oscillates about the minimum of its effective potential and ceases the acceleration promptly.  The onset of acceleration in our model is somehow similar to the symmetron model, but the presence of non-minimal couplings in teleparallel model provides a framework such that the quintessence can reside at the minimum of its effective potential giving rise to a de Sitter expansion ultimately. Hence our model, unlike the symmetron model, does not suffer from the short period of acceleration.

The scheme of the paper is as follows:
In the second section we introduce our model and, by obtaining scalar field solutions of the system, study $Z_2$ symmetry breaking which leads to an accelerated Universe after a decelerated matter dominated phase. In the third section, by employing a dynamical phase analysis,  we study the stability of late time solution corresponding to a de Sitter expansion.
In the fourth section, we discuss and conclude our results.

We will use units $\hbar=c=8\pi G=1$.

\section{The model, $Z_2$ Symmetry breaking and cosmic acceleration}
We start with the action
\begin{eqnarray}\label{8}
S&=&\int \left({T\over 2}+{1\over 2}\left(\partial_\mu \phi \partial^\mu \phi +\epsilon T \phi^2+\chi B \phi^2\right)-V(\phi)\right)\sqrt{-g}d^4x\nonumber \\
&+&\int \mathcal{L}_m(\tilde{e}^a_\mu)\sqrt{-\tilde{g}}d^4x
\end{eqnarray}
where
\begin{equation}\label{9}
\tilde{e}^a_\mu=A(\phi){e^a}_\mu,
\end{equation}
and $\tilde{g}_{\mu \nu}=A^2(\phi)g_{\mu \nu}$. The conformal coupling is given by the positive function $A(\phi)$. We consider a spatially flat FLRW space time  filled nearly with the scalar field dark energy and pressureless matter.  The scalar field equation of motion derived from (\ref{8}) is
\begin{equation}\label{10}
\ddot{\phi}+3H\dot{\phi}+V_{,\phi}-\left(\epsilon T+\chi B\right)\phi+A_{,\phi}A^{-1}\rho_m=0.
\end{equation}
The matter continuity equation reads
\begin{equation}\label{11}
\dot{\rho_m}+3H\rho_m=A_{,\phi}A^{-1}\rho_m\dot{\phi}.
\end{equation}
By defining the density
\begin{equation}\label{12}
\hat{\rho}_m=A^{-1}\rho_m,
\end{equation}
we find
\begin{eqnarray}\label{13}
&&\ddot{\phi}+3H\dot{\phi}+V_{,\phi}-\left(\epsilon T+\chi B\right)\phi+A_{,\phi}\hat{\rho}_m=0 \nonumber \\
&&{\dot{\hat{\rho}}_m}+3H\hat{\rho}_m=0.
\end{eqnarray}
Note that $\hat{\rho}_m$ is a mathematical $\phi$ independent variable, which facilitates our computations and interpretations.
In this space-time we have $T=-6H^2$ and $B=-18H^2-6\dot{H}$, where $H$ is the Hubble parameter which in terms of the scalar factor, $a(t)$, is given by $H={\dot{a}(t)\over a(t)}$.
Variation of (\ref{8}) with respect to the vierbeins yields
\begin{equation}\label{14}
H^2={1\over 3(1+\epsilon \phi^2)}\left({1\over 2}\dot{\phi}^2+V+6\chi H\phi \dot{\phi}+A\hat{\rho}_m\right),
\end{equation}
and
\begin{eqnarray}\label{15}
&&\dot{H}=-{1\over 2(1+(\epsilon+6\chi^2)\phi^2)}\Big((1-2\chi)\dot{\phi}^2+12\chi(\epsilon+3\chi)H^2\phi^2+\nonumber \\
&&4(\epsilon+3\chi)H\phi \dot{\phi}+2\chi \phi V_{,\phi}+A\hat{\rho}_m\Big).
\end{eqnarray}

Two of the three equations  (\ref{13}), (\ref{14}), and (\ref{15}) are independent. We consider (\ref{13}), and (\ref{14}) as independent equations. To solve them we must identify $V(\phi)$ and $A(\phi)$. As we want to study the $Z_2$ symmetry breaking, we choose the  potential as
\begin{equation}\label{16}
V(\phi)=-{1\over 2}\mu^2\phi^2+{\lambda\over 4}\phi^4,\,\,\, \lambda>0
\end{equation}
and take $A(\phi)$ as an even function of $\phi$, such that $A_{,\phi}\hat{\rho}_m$ in (\ref{10}) behaves as a $\hat{\rho}_m$ dependent mass term.
Based the coincidence problem and the theory of structures formation,  we expect that in the early Universe the contribution of dark energy was negligible. Therefore we assume that, initially
the scalar field stays at the minimum of its potential, identified by $\phi=0$, where $V(\phi=0)=0$ and $\dot{\phi}=0$ so
\begin{equation}\label{17}
\rho_\phi={1\over 2}\dot{\phi}^2+V(\phi)-3\epsilon H^2 \phi^2+6 \chi H \phi \dot{\phi}=0.
\end{equation}
In this epoch the Universe is dominated by matter evolving according to:
\begin{equation}\label{17.1}
\rho_m=A(\phi=0)\hat{\rho_m}(a=1)a^{-3}.
\end{equation}
Note that $A(\phi=0)$ is a nonzero number.  By using (\ref{13}), we introduce the scalar field effective potential,  $V_{eff}$, as
\begin{equation}\label{r1}
V_{eff,\phi}= V_{,\phi}-\left(\epsilon T+\chi B\right)\phi+A_{,\phi}\hat{\rho}_m.
\end{equation}
The scalar field stays in $\phi=0$, as long as its squared effective mass is still positive
\begin{equation}\label{18}
\mu_{eff}^2={\partial^2 V_{eff}\over \partial \phi^2}\big|_{\phi=0}>0.
\end{equation}
From (\ref{13}), and (\ref{r1}),  we find
\begin{equation}\label{19}
\mu_{eff}^2=-\mu^2+(2\epsilon+3\chi)\rho_m+A_{,\phi\phi}\big|_{\phi=0}\hat{\rho}_m.
\end{equation}
To build our model we need that $A_{,\phi\phi}\big|_{\phi=0}$ becomes a nonzero number. In this way, the sign of the effective squared mass depends on the value of the dark matter density. Like \cite{sym}, we take
\begin{equation}\label{20}
A(\phi)=1+{\phi^2\over 2 M^2},
\end{equation}
where $M$ is a mass scale. Hence (\ref{19}) becomes
\begin{equation}\label{21}
\mu_{eff}^2=-\mu^2+\left((2\epsilon+3\chi)+M^{-2}\right)\hat{\rho}_m.
\end{equation}
Thus provided that
\begin{equation}\label{22}
(2\epsilon+3\chi)+M^{-2}>0,
\end{equation}
as long as
\begin{equation}\label{23}
\hat{\rho}_m>{\mu^2 M^2\over (2\epsilon+3\chi)M^2+1}\equiv \rho_c,
\end{equation}
$\mu_{eff.}^2>0$ holds and the scalar field stays at the stable point $\phi=0$.  When $\hat{\rho}_m<\rho_c$, $\mu_{eff.}^2<0$ and this point becomes unstable. The scalar field becomes tachyonic and rolls to the new  minimum of its effective potential.  So in the first stage we have a Universe with zero dark energy density dominated by baryonic and cold dark matter.

In the second stage we require that dark energy density increases and gives rise to the Universe acceleration. To see how this expectation can be realized, let us obtain the minimum of the effective potential towards which the quintessence tends after the symmetry breaking. By bearing in mind that ${\hat \rho}_m={\hat \rho}_m(a=1)a^{-3}$, the late time solution corresponding to the minimum of the effective potential can be found from (\ref{13}) and (\ref{14}). This solution is characterized by $\dot{\phi}=0$, $\hat{\rho}_m= 0$ and
\begin{eqnarray}\label{24}
&&{V_{eff}}_{,\phi}=V_{,\phi}+6(\epsilon+3\chi)H^2\phi=0 \nonumber \\
&&H^2={1\over 3(1+\epsilon  \phi^2)}V.
\end{eqnarray}
 From (\ref{15}) we have also $\dot{H}=0$.  This solution corresponds to a de Sitter expansion with a constant $H$ and a constant $\phi$.

 Therefore our evolution equations correspond to General Relativity (GR) model with a cosmological constant and a modified gravitational constant specified in terms of $\phi$ in (\ref{24}). As shown in \cite{damour}, many classes of nonminimally coupled scalar theories have GR attractor solutions with a scalar field tending to a fixed value. In the GR limit, the Lorentz invariance holds. Thus in our model which possesses a late time GR attractor, the terms violating the Lorentz invariance and the spurious degrees of freedom are dynamically suppressed at the late time (for more discussion see  \cite{jar}).

 As said before, we can divide the evolution of our system into two eras. In the first stage, before the symmetry breaking,  the deceleration parameter $q=-1-{\dot{H}\over H^2}$, is  $q=0.5$.  In the second stage, as the late time critical point is given by $q=-1$,  we expect that Universe
enters a positive accelerated phase after matter dominated era. Finally, the system settles down in a de Sitter state.

Solutions of (\ref{24}),  when $\epsilon \neq -\chi$,  are given by
\begin{eqnarray}\label{25}
\phi_{\pm}^2&=&{3\chi \mu^2+2\epsilon \mu^2-\lambda\pm\sqrt{9\chi^2\mu^4+12\chi \epsilon \mu^4+4\epsilon^2\mu^4+2\epsilon\lambda\mu^2+\lambda^2}\over 3\lambda(\chi+\epsilon)}\nonumber \\
H_{\pm}^2&=&{\epsilon \mu^2+\lambda\mp\sqrt{9\chi^2\mu^4+12\chi \epsilon \mu^4+4\epsilon^2\mu^4+2\epsilon\lambda\mu^2+\lambda^2}\over 18(\chi+\epsilon)(3\chi +\epsilon)}.
\end{eqnarray}
These solutions are real provided that
\begin{eqnarray}\label{26}
&&(3\chi +2\epsilon)\mu^2-\lambda<0,\nonumber \\
&&9(\chi+{2\over 3}\epsilon)^2\mu^4+2\mu^2\epsilon \lambda+\lambda^2>0\nonumber \\
&&3\chi+\epsilon>0\nonumber \\
&&\epsilon\mu^2+\lambda<0\nonumber \\
&&\epsilon+\chi<0.
\end{eqnarray}
Also, when
\begin{eqnarray}\label{27}
&&(3\chi +2\epsilon)\mu^2-\lambda<0,\nonumber \\
&&9(\chi+{2\over 3}\epsilon)^2\mu^4+2\mu^2\epsilon \lambda+\lambda^2>0\nonumber \\
&&3\chi+\epsilon<0\nonumber \\
&&\epsilon+\chi<0,
\end{eqnarray}
only $\phi_{-}^2$ and $H_{-}^2$ exist.

The model reduces to the coupled scalar curvature model (\ref{5}) for $\epsilon =-\chi$. In this case (\ref{25}) is replaced with
\begin{eqnarray}\label{r2}
\phi^2&=&{\mu^2\over \epsilon\mu^2+\lambda}\nonumber \\
H^2&=&-{\mu^4\over 12(\epsilon\mu^2+\lambda)},
\end{eqnarray}
which shows that we cannot have real solutions and de Sitter solution does not exist. This is in agreement with the results of \cite{R1} which express that for scalar-curvature gravity the late time solution of the symmetron model, like the minimally coupled symmetron model,  is an oscillating (around the minimum of the effective potential )scalar field corresponding to a decelerating Universe. However having a final de Sitter expanding Universe may be accessible in models with a nonminimally coupled scalar field to the Ricci scalar (\ref{5}). Indeed in models other than the symmetron model and with potentials different from (\ref{16}), such as the exponential potential, the universe may experience a stable de Sitter state at late time \cite{R2}, or even crosses the phantom divide line \cite{R3}.

In the other limit, i.e.  $\chi =0$, the model reduces to the coupled scalar- torsion model (\ref{6}). Crossing the phantom divide line in this model was discussed in \cite{R4} for exponential potential, and in \cite{R5} for power law potentials. Our model with the potential (\ref{16}), as we will show in the last section, is capable of showing this characteristic. The existence of a de Sitter critical point at the late time and its stability were discussed in the literature (e.g. for power law potential see \cite{R6} and references therein). To obtain a power law potential we can set $\mu=0$ in (\ref{25}), resulting in
\begin{eqnarray}\label{r3}
\phi_{-}^2&=&-{2\over 3\epsilon}\nonumber \\
H_{-}^2&=&{\lambda\over 9\epsilon^2},
\end{eqnarray}
which is in agreement with \cite{R6}.  Note that for $\chi=0$, (\ref{26}) does not hold and only $\phi_{-}^2$ and $H_{-}^2$ exist. In this case, parameters of the model satisfy (\ref{27}).
So we conclude that by varying the parameters from the coupled scalar-curvature towards the coupled scalar-torsion, the de Sitter solution appears.

From (\ref{26}) or (\ref{27}) we can derive $2\epsilon+3\chi<0$. Whence, if we insist to have late time acceleration,  the initial stability  condition (\ref{22}) does not hold if we ignore the conformal coupling. In other words the conformal coupling
retains the  quintessence in the initial stable state for ${\hat{\rho}}>\rho_c$,  while the nonminimal couplings ($\chi,\,\,\,\epsilon$) are necessary for late time acceleration.
  To elucidate this subject let us see whether the same scenario can be set up in the minimal case, $\epsilon=\chi=0$. In the first stage when $\hat{\rho}_m>\mu^2 M^2$, the quintessence settles down in the minimum of its effective potential at $\phi=0$. When $\hat{\rho}_m<\mu^2 M^2$, the $Z_2$ symmetry breaks and the field moves towards the new minimum of the effective potential $\phi^2={\mu^2-{\hat{\rho}_m\over M^2}\over \lambda}$.   Finally, as $\hat{\rho}_m\to 0$, $\phi$ oscillates about $\phi=\sqrt{{\mu^2\over \lambda}}$. Note that as $V\left(\phi=\sqrt{\mu^2\over \lambda}\right)=-{\mu^2\over 4\lambda}<0$, the solution (\ref{24}), which gives a negative $H^2$, is not valid in this case and the field cannot stay at the minimum of its potential but oscillates about it and acts as a pressureless matter \cite{sami},  giving rise to a decelerated expansion. Moreover, If one considers local gravity tests to screen the new force mediated by the scalar field, he obtains a large mass for the scalar field which expedites this overshoot. Besides, after the symmetry breaking the quintessence potential (\ref{16}) becomes negative and as it is known a quintessence with negative potential cannot drive the acceleration in the minimal case \cite{sadjadi1}. To remedy this, some authors add a positive constant to (\ref{16}) playing the role of a cosmological constant rendering the problem to $\Lambda CDM$ scenario.
\section{Phase space analysis}

Phase space analysis may be employed to investigate the stability of the solution (\ref{25}). To do so we define
\begin{equation}\label{28}
x={{\dot \phi}\over \sqrt{6}H},\,\, y=\phi,\,\,u={\sqrt{\hat{\rho}_m}\over \sqrt{3} H}.
\end{equation}
these variables satisfy the autonomous equations
\begin{eqnarray}\label{29}
&&x'=E(x,y,u)=-3x-\sqrt{6}(\epsilon+\chi(3+s))y-sx-{3\over \sqrt{6}}f(y)(1-u^2-x^2\nonumber \\
&&-2\sqrt{6}\chi xy+\epsilon y^2)\nonumber \\
&&y'=F(x,y,u)=\sqrt{6}x\nonumber \\
&&u'=G(x,y,u)=-{3\over 2}u-su,
\end{eqnarray}
where $f(y)={V_{,\phi}\over V}$ and prime denotes derivative with respect to $\ln(a)$, and $s={\dot{H}\over H^2}$ is given by
\begin{eqnarray}\label{30}
&&s(x,u,y)=-\left({1\over 1+(6\chi^2+\epsilon)y^2}\right)\Big({3\over 2}u^2+3(1-2\chi)x^2+6\chi(\epsilon+3\chi)y^2\nonumber \\
&&+2\sqrt{6}(\epsilon+3\chi)xy+ 3\chi y f(y)(1-u^2-x^2-2\sqrt{6}\chi x y+\epsilon y^2)\Big).
\end{eqnarray}
To derive (\ref{30}) we have used the Friedmann equation
\begin{equation}\label{31}
u^2+x^2+{V\over 3H^2}+2\sqrt{6}\chi xy-\epsilon y^2=1.
\end{equation}
The critical point corresponding to the late time evolution of the Universe is given by $\{x=0,y=y_c,u=0,s=0,\}$ corresponding to (\ref{25}), where $y_c=\phi_{\pm}$.
The stability of this point may be checked as follows. Setting $x\to x+\delta x$,  $y\to x+\delta y$, $u\to u+\delta u$, we find
\begin{equation}\label{32}
{d\over d{\ln a}}{ \left( \begin{array}{cccc}
\delta x\\
\delta y \\
\delta u \\
\end{array} \right)}=\mathcal{N} \left( \begin{array}{cccc}
\delta x\\
\delta y \\
\delta u \\
\end{array} \right),
\end{equation}
where
\begin{equation}\label{33}
\mathcal{N}=\left( \begin {array}{cccc} E_{,x}& E_{,y} &E_{,u}  \\ \noalign{\medskip}F_{,x}&F_{,y}&F_{,u}
\\ \noalign{\medskip}G_{,x}&G_{,y}&G_{,u}\end {array} \right).
\end{equation}
At the critical point $\{x=0,u=0,s=0,y=y_c\}$, $\mathcal{N}$ becomes
\begin{equation}\label{34}
\mathcal{N}_c=\left( \begin {array}{cccc} N_{11}& N_{12} &0 \\ \noalign{\medskip}\sqrt{6}&0&0
\\ \noalign{\medskip}0&0&-{3\over 2}\end {array} \right),
\end{equation}
where
\begin{eqnarray}\label{35}
&&N_{11}=-3-\sqrt{6}\chi y_c s_{,x}(0,y_c,0)+6\chi y_cf(y_c)\nonumber \\
&&N_{12}=-\sqrt{6}\chi y_c s_{,y}(0,y,0)-\sqrt{6}(\epsilon+3\chi)\nonumber \\
&&-\sqrt{3\over 2}f_{,y}(y_c)(\epsilon y_c^2+1)-\sqrt{6}\epsilon y_c f(y_c).
\end{eqnarray}
From (\ref{29}), we obtain
\begin{equation}\label{36}
f(y_c)=-{2(\epsilon+3\chi)y_c\over 1+\epsilon y_c^2}
\end{equation}
Inserting (\ref{36}) into (\ref{35}), and by considering the potential (\ref{16}),  we find
\begin{equation}\label{37}
N_{{12}}=3\sqrt{6}{P(y_c)\over {y_c^{2} \left( \lambda\,{y_c}^{2}-2\,{\mu}^{2} \right) ^{2} \left( 6\,{
\chi}^{2}{y_c}^{2}+\epsilon\,{y_c}^{2}+1 \right)}
},
\end{equation}
where
\begin{eqnarray}\label{38}
P(y_c)&=&{\lambda}^{2}\epsilon\, \left( \chi+\epsilon \right) {y_c}^{8}-4\,
 \left(  \left( \chi/4-\epsilon/4 \right) \lambda+{\mu}^{2}\epsilon\,
 \left( \chi+\epsilon/2 \right)  \right) \lambda\,{y_c}^6+\nonumber \\
 &&\left( 2/3\,{\lambda}^{2}+4\,\chi\,\lambda\,{\mu}^{2}+4\,{\mu}^{4}
\epsilon\, \left( \chi+2/3\,\epsilon \right)  \right) {y_c}^{4}-\nonumber \\
&&4\,{\mu}
^{2} \left( \lambda/6+{\mu}^{2} \left( \chi-\epsilon/3 \right)
 \right) {y_c}^{2}+ 4/3\,{\mu}^{4}.
\end{eqnarray}
So eigenvalues of (\ref{34}) are
\begin{equation}\label{39}
-{3\over 2},\,\,\, -{3\over 2}+{1\over 2}\sqrt{9+4\sqrt{6}N_{12}},\,\,\, -{3\over 2}-{1\over 2}\sqrt{9+4\sqrt{6}N_{12}}.
\end{equation}
If we require that all eigenvalues to be negative, which is sufficient condition to have an attractor, we must take
\begin{equation}\label{40}
N_{12}<0.
\end{equation}
By using (\ref{37}) and (\ref{40}) we obtain
\begin{eqnarray}\label{41}
{P(y_c)\over { 6\,{
\chi}^{2}{y_c}^{2}+\epsilon\,{y_c}^{2}+1}}<0,
\end{eqnarray}
which is the condition to have an attractor at late time.

We remember that for $\epsilon=-\chi$, we have not a late time de Sitter attractor (see the discussion after (\ref{r2})). The  presence of $\chi\neq 0$ provides more possibilities to have stable solution, for example in a model with $\mu=0$ which corresponds to the models discussed in \cite{R6}, the condition (\ref{40}) becomes ${\epsilon\over 36 \chi^2+9\chi+15\epsilon}<0$, which does not hold when $\chi=0$.

\begin{figure}[H]
\centering\epsfig{file=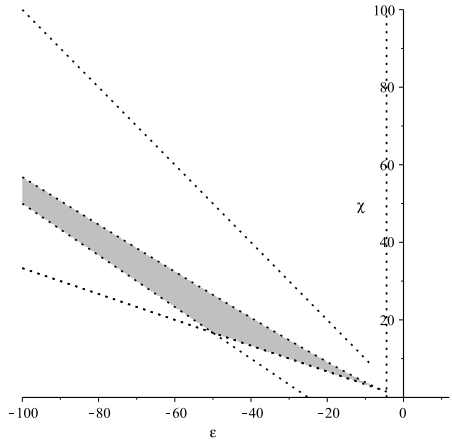,width=6cm,angle=0}
\caption{ An illustration of allowed domain (grey color) for $\chi$ and $\epsilon$ for ${\lambda \over H_0^2} =10^3,{\mu\over H_0}=15, {1\over M^2}=50$. } \label{fig1}
\end{figure}

To get an intuition about the range of the parameters, let us depict the domain of validity of (\ref{22}), (\ref{26}), and (\ref{41}) in terms of  $\epsilon$ and $\chi$ for $\{{\lambda \over H_0^2} =10^3,{\mu\over H_0}=15, {1\over M^2}=50\}$ in Figure (\ref{fig1}). The allowed domain is shown by grey color.

\section{Discussions and Conclusion}
\begin{figure}[H]
\centering\epsfig{file=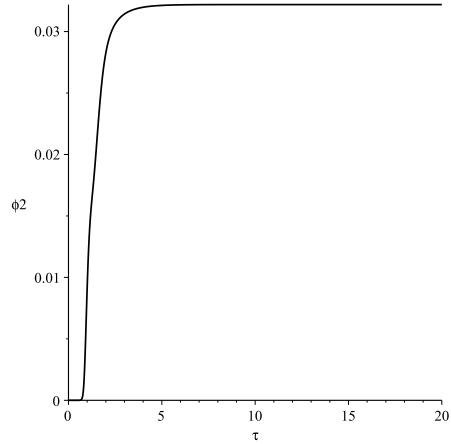,width=6cm,angle=0}
\caption{ An illustration of $\phi^2$  in terms of $\tau=tH_0$ for $\{{\lambda\over H_0^2} =10^3,{\mu\over H_0}=15, \chi=10, \epsilon=-29,{1\over M^2}=50\}$ and with the initial conditions
$\{\phi(0)=0,{d{\phi}\over d\tau}(0)=0.0001,{\rho_{m}(0)\over H_0^2}=9\}$.} \label{fig2}
\end{figure}

\begin{figure}[H]
\centering\epsfig{file=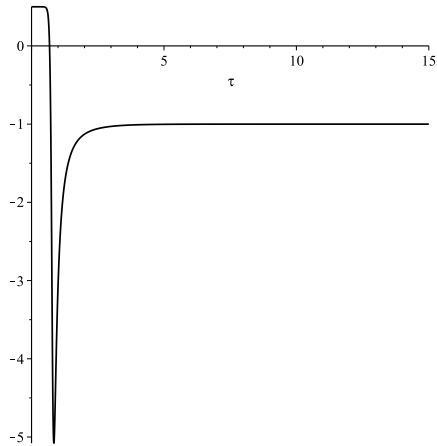,width=6cm,angle=0}
\caption{ An illustration of deceleration  factor in terms of $\tau=tH_0$ for $\{{\lambda\over H_0^2} =10^3,{\mu\over H_0}=15, \chi=10, \epsilon=-29,{1\over M^2}=50\}$ and with the initial conditions
$\{\phi(0)=0,{d{\phi}\over d\tau}(0)=0.0001,{\rho_{m}(0)\over H_0^2}=9\}$.} \label{fig3}
\end{figure}

We tried to attribute the late time cosmic acceleration to symmetry breaking in a teleparallel model of cosmology in which quintessence is coupled no-minimally to both the scalar torsion and a boundary term. We derived equations of motion, from which suitable solutions were derived. The solutions were classified into two classes: before symmetry breaking and after symmetry breaking. Before symmetry breaking, provided that (\ref{22}) holds, the conformal coupling holds the quintessence in the bottom of its effective potential. In this epoch dark energy has no contribution in total energy, and so the coincidence problem may be alleviated in this model. When the density of matter becomes less than a critical value (given by (\ref{23})), the initial state becomes unstable and the quintessence moves towards the minimum of the effective potential which describes a state with de Sitter expansion (provided that the conditions (\ref{26}) or (\ref{27}) be satisfied). Finally, if (\ref{41}) holds, the scalar filed resides at this stable state  due to the presence of the nonminimal couplings. In figures (\ref{fig2}) and (\ref{fig3}), the behaviors of the quintessence and the deceleration parameter are depicted in terms of the dimensionless time $\tau=tH_0$ for $\{{\lambda\over H_0^2} =10^3,{\mu\over H_0}=15, \chi=10, \epsilon=-29,{1\over M^2}=50\}$ and with the initial conditions $\{\phi(0)=0,{d{\phi}\over d\tau}(0)=0.0001,{\rho_{m}(0)\over H_0^2}=9\}$. If we reset the natural units $\hbar=c=1$, these conditions become $\{{\lambda M_P^2\over H_0^2} =10^3,{\mu\over H_0}=15, \chi=10, \epsilon=-29,{M_P^2\over M^2}=50, \}$ and $\{{\phi(0)\over M_P}=0,{1\over M_P}{d{\phi}\over d\tau}(0)=0.0001,{\rho_{m}(0)\over H_0^2M_P^2}=9\}$, where $M_P=\sqrt{1\over 8\pi G}$ is the reduced Planck mass. Figure (\ref{fig2}) shows that the scalar field leaves $\phi=0$ after the symmetry breaking and tends towards the late time critical point and finally settles down in it. Figure (\ref{fig3}) shows that, during this evolution,  the deceleration factor becomes negative in a time of order of the present Hubble time. This figure also shows the possibility to have super-acceleration phase $q<-1 (\dot{H}>0)$ which is forbidden in the minimal case (see(\ref{15})). This crossing was also reported in \cite{R4} and \cite{R5}, for exponential and power law potentials.


\end{document}